\documentclass[11pt,a4paper]{article}
\usepackage{ifpdf}
\usepackage{ae}
\usepackage[T1]{fontenc}
\usepackage[ansinew]{inputenc}
\usepackage{mathrsfs}
\usepackage{amsmath}
\usepackage{amssymb}
\pdfoutput=1
\usepackage{amsmath,amssymb,amsfonts,a4wide,graphicx,bm,times,psfrag,wrapfig,sidecap}
\usepackage{cite}
\usepackage[colorlinks=true,linkcolor=black, citecolor=black,
urlcolor=black]{hyperref} 
\numberwithin{equation}{section}
\makeatletter \let\old@startsection=\@startsection
\renewcommand{\@startsection}[6]
{\old@startsection{#1}{#2}{#3}{#4}{#5}{#6\mathversion{bold}}}
\makeatother
\def\O{\Omega}

\def\defeq{\stackrel{\text{def}}=}
\newcommand\re[1]{({\ref{#1}})}
\def\be{\begin{eqnarray}  }
    \def\ee{\end{eqnarray}}

\def\IZ{{\mathbb{Z}}}

    \def\no{\nonumber}
    \def\la{\label}

    \def\({\left(} \def\){\right)} \def\<{\left\langle\,} \def\>{\,
    \right\rangle} \def\[{\left[} \def\]{\right]} 
    \def\hf{ {\textstyle{1\over 2}} }  \def\CA{{\cal A}} \def\CD{{\cal D}} 
    \def\bS{{\mathblackboard S}}  
     \def\CZ{{ \mathcal{ Z} }}  
     \def\CC{ {\mathcal C}}  \def\CN{{ \cal N}}     
     \def\p{\partial}  
       
     \def\th{\theta}  
 \def\Tr{{\rm Tr}}

\def\d{\delta}
\newcommand{\caS}{{\mathscr S}}
\newcommand{\caA}{{\mathscr A}}

\newcommand{\caZ}{{\mathscr Z}}

\newcommand\braup[1]{\, \langle \uparrow ^{ #1}\!\! |}
\newcommand\ketup[1]{|  \! \uparrow ^{ #1} \rangle}
\newcommand\bradown[1]{\langle \, \downarrow^{#1}\!\! | }
\newcommand\ketdown[1]{|  \! \downarrow^{#1} \rangle }

 \newcommand\spinup[1]{\uparrow\uparrow.
..^{\!\!\!  \!\!  #1}\uparrow }
 \newcommand\Dp[1]{ e^{i\p/\p {#1}}} \newcommand\Dm[1]{
e^{-i\p/\p {#1}}}  
 \def\ket{ | 0 \rangle} \def\bra{ \langle 0
| }

  \def\bT{{\mathcal T}} \def\bS{{\mathbf S}}
 \def\bR{{\mathcal R}}

  \usepackage{bm}% bold math
\def\O{\Omega}
\def\defeq{\stackrel{\text{def}}=}
\def\ee{\end{eqnarray}}  
\def\IZ{{\mathbb{Z}}}  
    \def\no{\nonumber} \def\la{\label} 
\def\({\left(} \def\){\right)} \def\<{\left\langle\,} \def\>{\,
\right\rangle} \def\[{\left[} \def\]{\right]} 
\def\hf{ {\textstyle{1\over 2}} } 
\def\CA{{\mathcal A}}  \def\CB{{\cal B}}   \def\CZ{{ \mathcal{ Z} }}
 
 \def\CC{ {\mathcal C}}  \def\CN{{ \cal N}}  
    \def\p{\partial}   
   
\def\th{\theta}   \def\Tr{{\rm Tr}}
 \def\d{\delta}  \def\zz{
{ { \bf z} }} \def\uu{ { {\bf u} }} \def\vv{ { \bf v}} \def\ww{ { \bf
w }}  \def\vv { { \bf v }}   
\def\bS{{\mathbb S}}
\def\llangle{   \langle  }
\def\rrangle{  \rangle }
\def\vvert{\vert   }

\newcommand\encadremath[1]{\vbox{\hrule\hbox{\vrule\kern8pt
\vbox{\kern8pt \hbox{$\displaystyle #1$}\kern8pt}
\kern8pt\vrule}\hrule}} \def\enca#1{\vbox{\hrule\hbox{
\vrule\kern8pt\vbox{\kern8pt \hbox{$\displaystyle #1$} \kern8pt}
\kern8pt\vrule}\hrule}}

\begin{document}

\thispagestyle{empty}

\begin{flushright}
  IPhT/t12/049  \\
  UT12-16
\end{flushright}

\vspace{1cm}
\setcounter{footnote}{0}

\begin{center}

 {\Large\bf Inner products of Bethe states as  partial domain wall
 partition functions }

\vspace{20mm} 

Ivan Kostov$^*$\footnote{\it Associate member of the Institute for
Nuclear Research and Nuclear Energy, Bulgarian Academy of Sciences, 72
Tsarigradsko Chauss\'ee, 1784 Sofia, Bulgaria} and Yutaka
Matsuo$^\dag$ \\[7mm]
 
{$^*$\it Institut de Physique Th\'eorique, CNRS-URA 2306 \\
	     C.E.A.-Saclay, \\
	     F-91191 Gif-sur-Yvette, France} \\[5mm]

{$^\dag$\it Department of Physics, The University of Tokyo\\
Hongo 7-3-1, Bunkyo-ku\\
Tokyo 113-0033, Japan} \\[5mm]

\end{center}

\vskip9mm

\vskip18mm

 \noindent{ We study the inner product of Bethe states in the
 inhomogeneous periodic XXX spin-1/2 chain of length $L$, which is
 given by the Slavnov determinant formula.  We show that the inner
 product of an on-shell $M$-magnon state with a generic $M$-magnon
 state is given by the same expression as the inner product of a
 $2M$-magnon state with a vacuum descendent.  The second inner product
 is proportional to the partition function of the six-vertex model on
 a rectangular $L\times 2M$ grid, with partial domain-wall boundary
 conditions.
 
 % Lattice Integrable Models, Bethe Ansatz, 
 
  %   
\newpage
\setcounter{footnote}{0}

%&&&&&&&&&&&&&&&&&&&&&&&&&&&&&&&&&
\section{Introduction}
%&&&&&&&&&&&&&&&&&&&&&&&&&&&&&&&&&
\label{sec:Intro}

The computation of the inner product of two Bethe wave functions is a
necessary step on the way of obtaining the correlation functions in
integrable models.  An expansion formula for the scalar product of two
generic Bethe states in the generalized $su(2)$ model is due to
Korepin \cite{korepin-DWBC}.  The sum formula becomes increasingly
difficult to tackle when the number of pseudoparticles becomes large.
Simplifications occur when one of the states is on shell, {\it i.e.},
when its rapidities satisfy the Bethe equations.  In this case the
inner product can be cast in the form of a determinant.  A determinant
formula for the norm of an on shell Bethe state was originally
conjectured by Gaudin \cite{Gaudin-livre} and later proved by Korepin
\cite{korepin-DWBC}.  A determinant formula for the scalar product of
one on shell and one generic Bethe states was found by N. Slavnov
\cite{NSlavnov1}.\footnote{ Previously, a representation of the inner
product as a determinant of double size was found by Kirillov and
Smirnov \cite{KirSm}.} The Slavnov formula was used to obtain some
correlation functions in the XXX and XXZ spin chains
\cite{2005hep.th....5006K}, and more recently in the computation of 
some correlation functions in the $\CN=4$ supersymmetric Yang-Mills
theory \cite{EGSV, GSV, 3pf-prl, SL }

If the two Bethe states are composed from $M$ pseudoparticles with
rapidities $\uu=\{u_1,\dots, u_M\}$ and $\vv= \{v_1,\dots, v_M\}$, the
Slavnov inner product is given, up to a simple factor, by a $M\times
M$ determinant.  In \cite{3pf-prl, SL} one of the authors derived a a
closed expression of the inner product for the XXX spin chain in the
classical limit $M\to\infty$, in which the Bethe roots condense into
several macroscopic strings.\footnote{The classical limit was studied
in \cite{Kazakov:2004qf}; in the condensed matter physics it is known
also as Sutherland limit \cite{PhysRevLett.74.816}.  } An intriguing
feature of this expression is that it has the same form as the
classical limit of the scalar product of a Bethe state with a vacuum
descendent, obtained previously in \cite{GSV}, if one chooses the
rapidities $\ww$ of the Bethe state as $\ww=\uu\cup \vv$.

 In this paper we prove that the above is true also for finite $M$.
 We show that the Slavnov inner product $\langle \vv|\uu\rangle$ of
 two $M$-magnon states in a periodic inhomogeneous XXX chain with spin
 $1/2$ is equal to the inner product of a vacuum descendent with the
 $2M$-magnon state $|\uu\cup \vv\rangle$, which is in turn
 proportional to the partition function of the six-vertex model with
 partial domain wall boundary conditions, studied recently in
 \cite{FW3}.

\section{ Inner product of Bethe states in the inhomogeneous XXX 1/2
chain } \la{sect:Rappel}

%\subsection{The space of states as a Fock space}

The XXX spin chain is characterized by an $R$-matrix $R_{12}(u,v)$
acting in the tensor product $V_1\otimes V_2$ of two copies of the
target space.  Up to a scalar factor, the $R$-matrix of the XXX spin
chain is \cite{Fad-Tacht-84}
  \be \la{defR} R_{12}(u,v) = u-v+ {iP_{12} }\, , \ee
where $P_{12} $ is the permutation operator acting in the tensor
product $V_1\otimes V_2$ of two copies of the target space.  The
inhomogeneous XXX spin chain of length $L$ is characterized by
background parameters (impurities) $\zz=\{ z_1,\dots , z_L\}$
associated with the $L$ sites of the chain.  For the spin 1/2 chain,
the monodromy matrix $\bT_a (u)\in \text{End}(V_a)$ represents the
product of the $R$-matrices along the spin chain,
    \be \la{defTa} \bT_a(u) \equiv R_{a1}(u, z_1 )\, R_{a2}(u, z_2
    )\dots R_{aL}(u,z_L ) = \begin{pmatrix} {\CA}(u) &{ \CB}(u) \\
    \CC(u)  &  \CD(u)
    \end{pmatrix} .
 \ee
The homogeneous XXX spin chain corresponds to the limit $z_m\to i/2
$.\footnote{ Sometimes $\th_j= z_j- i/2$ are referred to as
inhomogeneity parameters.  With this definitioon the homogeneous limit
corresponds to $\th_j=0$.}

The matrix elements $\CA,\CB,\CC, \CD$ are operators in the Hilbert
space $V= V_1\otimes\dots \times V_L$ of the spin chain.  The commutation
relations between the elements of the monodromy matrix are determined
by the RTT relation
  \be\la{RTT} R_{12}(u-v) \bT_1(u) \bT_2(v) = \bT_2(v) \bT_1(u) \,
  R_{12}(u-v) ,
  \ee
 which follows from the Yang-Baxter equation for $R$.  
 In components,
   \be\la{TRRCC}
   \begin{aligned}
    \CA(v) \CB(u) &= \textstyle{u-v+i\over u-v} \ \CB(u) \CA(v)
    -\textstyle{i\over u-v} \CB(v) \CA(u), \\
   \CD(v) \CB(u) &= \textstyle{u-v-i\over u-v} \ \CB(u) \CD(v) 
   +  \textstyle{i\over u-v} \CB(v) \CD(u),
   \\
   [\CC(u), \CB(v)] &= \textstyle{i\over u-v} \( \CA(v) \CD(u) -
   \CA(u)\CD(v)\), \quad \text{etc}.
   \end{aligned}
   \ee
 As a consequence of \re{RTT}, the families of operators $\CB(u)$,
 $\CC(u)$, as well as  the transfer matrices  
  \be \la{defTrM} 
  \bT (u) \equiv\Tr_a [  \bT_a(u)] =   \CA(u) +
    \CD(u) \, ,
  \ee
  are commuting.
 
In the Algebraic Bethe Ansatz, the Hilbert space has the structure of
a Fock space generated by the action of the creation operators
$\CB(u)$ on the pseudo-vacuum $ \ketup{L}= \big|\, \spinup{L }
\big\rangle $, where all spins are oriented up.  The pseudo-vacuum is
an eigenstate for the diagonal elements $\CA$ and $\CD$ and is
annihilated by $\CC$:
 \be \la{acket}
 \( \CA(u)- A(u)\)
 \ \ketup{L}=  
 \( \CD(u)- D(u)\)\ketup{L} = \CC(u)\ketup{L}= 0,
   \ee
 where
 \be
 \la{defad}
 \begin{aligned}
 A(u) = \prod_{m=1}^L \(u- z_m+i\) , \quad D(u) = \prod_{m=1}^L\( u-
 z_m\) .
\end{aligned}
 \ee
 A creation/annihilation
 operator with $u\to\infty$ is the global $su(2)$ lowering/raising
 operator,
\be
 {\CB(u)\over D(u) } \simeq {i\over u} \bS^-,
 \quad
  {\CC(u)\over A(u) } \simeq {i\over u} \bS^+
 . 
 \ee

 The dual Bethe states are generated by the action of the
 $\CC$-operators on the dual pseudo-vacuum $\braup{L}= \big\langle
 \spinup{L } \big|$, which is annihilated by the $\CB$-operators,
% %
 and the hermitian conjugation can be defined as
 \be \CC(u) = - \CB^\dag( \bar u).  \ee
The space of states is a closure of the linear span of all vectors of
the form
\be \la{genstateket} \vvert \uu \rrangle = \prod_{j=1}^{M} \CB(u_j )\
\ketup{L} , \qquad \uu = \{u_1,\dots u_{_M}\}  , \ \ M\equiv \#\uu.
  \ee
The operator $\CB(u)$ can be viewed as a creation operator of a
pseudoparticle (magnon) with momentum $p= \log{u+i/2\over u-i/2}$.
Such states are called generic, or off shell, Bethe states.  The
scalar product of two generic Bethe states,\footnote{ Here we abuse
slightly the established notations, since with the convention
$\CB(u)^\dag = - \CC(\bar u)$ , the state dual to $|\vv\rangle $ is
$(-1)^N \langle \bar \vv|$.  In our notations the norm of a Bethe state
is $||\uu || ^2 = (-1)^N \langle \bar \uu|\uu \rangle$.  If one of the
states is on shell, then the set of its rapidities is invariant upon
complex conjugation, and the inner products $(-1)^N \langle \bar \vv|\uu
\rangle$ and $ \langle \vv|\uu \rangle$ differ only by a phase factor.
}

\be \la{FockSL} \langle \uu | \vv\rangle\, = \braup{L} \prod_{k=1}^{M}
\CC(v_k ) \prod_{j=1}^{M} \CB(u_k ) \ketup{L} \, , \ee
can be computed by applying the relations of the RTT algebra \re{RTT}.
For example, the scalar product of two one-magnon states is
\be \langle v|u\rangle = {i\over u-v}\( A(v) D(u) - A(u)D(v)\).
\la{onemagnongen} \ee

A Bethe state is an eigenvector of the transfer matrices [Eq.
\re{defTrM}] if the rapidities $\uu= \{u_1,\dots u_M\}$ satisfies the
{\it on-shell condition}, which is given by the Bethe equations
\cite{Fad-Tacht-84}
\be \la{BAEXXX} \prod_{k =1}^M {u_j - u_k +i\over u_j -u_k - i} =-
\prod_{m=1}^L {u_j-z_m + i \over u_j-z_m}, \quad j=1, \dots, M. \ee
To avoid lengthy formulas, throughout this paper we will use
systematically the following notations.  For any set $\ww=
\{w_j\}_{j=1}^N$ of points in the complex plane, we define the Baxter
polynomial
\be \la{defQ} Q_\ww(u) \defeq \prod_{j =1}^N (u- w_j ) , 
\quad N = \#\ww,
\ee
as well as the rational  function
\be \la{defEpm} E^\pm_\uu(u) \defeq {Q_\uu(u\pm i)\over Q_\uu(u)}.
\ee
In these notations, the eigenvalue $T_\uu (u)$ of the transfer matrix
on the on-shell state $|\uu\rangle$ is
    \be
    \begin{aligned}
     \la{eigenvalueT} T_{\uu}(u) 
    &= Q_\zz(u+i)\  E^-_{\uu}(u)  + Q_\zz(u)\   E^+_{\uu}(u).
    \end{aligned}
     \ee
  Another way to write the Bethe equations  is as
 \be \la{BAE} e^{2ip_\uu(u)}=-1, \quad u\in \uu,   \ee
 where  the  pseudomomentum $p_\uu$, known also as 
 counting function, is defined
 modulo $\pi$ by 

\be \la{defquazimomentum} e^{2i p_\uu } = {1\over E^+_\zz}\
{E^{+}_\uu\over E^-_\uu} \, .  \ee

As shown by Slavnov \cite{NSlavnov1}, when the state
$\vvert\uu\rrangle$ is on shell, the inner product with a generic
Bethe state $\llangle \vv \vvert$ is a determinant.  One can write the
Slavnov determinant formula as
   \be \la{defSuv0} \llangle \vv \vvert \uu\rrangle &=&\prod_{j =1}^M
   A(v_j ) D(u_j )\ \caS_{\uu, \vv}\, ,  
   \la{defscpr} 
   \\
   \caS_{\uu, \vv} &=& {\det_{j k}\O(u_j , v_k ) \over \det_{j k}
 {1\over u_j -v_k +i} } 
 \la{defSuv}\, ,  
 \ee
where the Slavnov kernel $\O(u,v)$ is defined by
  \be 
  \la{defOhat} \O(u,v)= t(u-v) - e^{2 i p_\uu(v)} \ t(v-u)\, ,
  \qquad t(u) = {1\over u } - {1\over u+i}\, .  
  \ee
  For example, Eq.  \re{defSuv} gives for the inner product of two
  one-magnon states $(M=1)$
 \be \la{SL1} \langle v|u\rangle = A(v)D(u) \, \caS_{u,v},\quad
 \caS_{u,v} = i {1- {D(v)\over A(v) }\over u-v}.  \ee
This expression indeed matches with the restriction of the general
expression \re{onemagnongen} when the rapidity $u$ is taken on shell,
$D(u)/ A(u)=1$.

   \section{Alternative expression for the inner product }
 \la{sec:operfac}

\subsection{ Operator factorization formulas}
 \la{sec:operfactor}

The Slavnov determinant \re{defSuv} can be given a very convenient
operator expression \cite{3pf-prl, SL}, whose derivation we review
below.  We represent the Slavnov kernel $\O(u,v)$ as the result of the
action of two difference operators on the Cauchy kernel ${1\over
u-v+i}$,
\be \la{SlM1} \O(u,v)&=& (1- e^{2ip_\uu(v)} \Dp{v})\, (\Dm{u} -1) \
{1\over u-v+i}\, , \ee
and write the Slavnov determinant as the result of the action of $N$
pairs of difference operators to the Cauchy determinant ,
\be \la{findifex} \caS _{\uu,\vv} ={ \prod_{v\in\vv} \(1 -e^{2ip_{_\uu}
(v)} e^{i\p/\p v} \)\prod_{u\in\uu} \( e^{-i \p/\p u} -1 \)\, \det_{j k}
{1\over u_j -v_k +i} \over \det_{j k} {1\over u_j -v_k +i} }\ .  \ee
Here, and in the following, the formulae contain products of
difference operators and the ordering of the difference operators
should be respected.  The factors within each of the two blocks in the
above formula commute, but the factors belonging to different blocks
do not.

 Now we apply the Cauchy identity
 \be \det_{j k} {1\over u_j -v_k +i}= { \prod_{j<k} (u_j-u_k) \
 \prod_{j<k}(v_k-v_j) \over \prod_{j , k =1}^M (u_j -v_k +i )} \equiv
 {\Delta_\uu\ \Delta_{-\vv} \over \Pi_{\uu, \vv}}\, .  \ee
 After repeated application of the obvious identities
 \be \la{CommD} e^{-i \p/\p u} \ 
{1\over  \Pi_{\uu, \vv}}
 &=& { E^+_\vv(u) }  \ {1\over  \Pi_{\uu, \vv}}\ e^{-i
 \p/\p u}  \qquad (u\in \uu) \no \\
  e^{i \p/\p v}  \ {1\over \Pi_{\uu, \vv}}\ &=& {
  E^-_{\uu} (v_j ) } \ {1\over  \Pi_{\uu, \vv} }\ e^{i
  \p/\p v_j}
  \qquad (v\in\vv)
  \ee
and taking into account the expression \re{defquazimomentum} for the
pseudomomentum, we write Eq.  \re{findifex} in a factorized operator
form,
  \be \la{FactorFb} \caS_{\uu,\vv} =(-1)^{\# \uu}\ { 1 \over
  \Delta_\vv } \prod_{v\in\vv}\(1 - {E^+_\uu(v)\over E^+_\zz(v)} \,
  e^{i \p/\p {v}} \) \Delta_\vv \ \cdot { 1 \over \Delta_{\uu} }
  \prod_{u\in\uu} \(1 - E_\vv^+ (u) \, e^{- i\p/\p{u}} \) \Delta_{\uu}
  .  \ee 
Here, we have to be careful in that the operator
$\exp(i\partial/\partial v)$ acts on all factors on the right of it.

  \subsection{The functional $\caA_\uu[f]$}
  
The two blocks of factors in the above operator expression have a
similar form and suggest introducing the following quantity.  For any
set of points $\uu= \{u_j \}_{j =1}^M$ in the complex plane and for
any complex function $f(z)$, we define the functional
\be \la{defCA}
% \encadremath{ 
 \begin{aligned}
 \caA^\pm _\uu[f] &\defeq { 1 \over \Delta_{\uu} } \prod_{u\in \uu}
 \(1 - f(u ) \, e^{\pm i \p/\p u } \) \Delta_{\uu}\,.
\end{aligned}
%}
 \ee
 Substituting $\Delta_\uu= \det_{jk} (u_j^{k-1})$, one can write this
 functional as a ratio of determinants \cite{3pf-prl, SL} \be
 \caA^\pm _\uu[f] &=& { \det_{j k} \( u_j ^{k-1} - f(u_j ) \, (u_j \pm
 i)^{k-1}\) \over \det_{j k} \( u_j ^{k-1}\) } \, , \ee
but for our purposes the operator representation \re{defCA} is more
convenient.
 
The functional $\caA^\pm_\uu[f]$ can be expanded as a sum of monomials
associated with the partitions of the set $\uu$ into two disjoint
subsets,
\be \la{ExpA} 
\begin{aligned}
\caA_{\uu}^\pm [f]&= \sum_{\uu'\cup \uu''=\uu }
\caA^\pm_{\uu', \uu''} \prod_{u''\in \uu''} [-f(u'')]\, , 
\qquad
\caA_{\uu', \uu''}^\pm
= \prod _{ u''\in\uu''} {E^\pm_{\uu'}(u'') }
.
\end{aligned}
 \ee
Under this form, the functional $\caA^\pm_\uu[f]$ appeared previously in
Ref.  \cite{GSV}.  If the function $f(u)$ depends implicitly on $\uu$,
we define the functional $\caA^\pm_\uu[f]$ so that it is given by the same
expansion \re{ExpA}.

The operators $ \caA^+ _\uu[f]$ and $ \caA^- _\uu[f]$ are related by
the functional identities
   \be
   \la{funceqa} 
   \caA^- _{\uu}[f]&=&
   \caA^+_{\uu}[1/f] \ \prod_{u\in\uu } [- f(u )] \, , 
   \quad
    \caA^+ _{\uu}[f]=
   \caA^-_{\uu}[1/f] \ \prod_{u\in\uu } [- f(u )] \, , 
  \\
   \la{funceqb}
    \caA^- _{\uu}[f]&=& \caA^+_\uu \big[ -
   {E^-_\uu\over E^+_\uu} \, f\big] \, ,
   \hskip 1.5cm  \caA^+ _{\uu}[f]=
   \caA^-_\uu \big[ - {E^+_\uu\over E^-_\uu} \, f\big] \ . 
   \ee
The first pair of identities was proved in \cite{3pf-prl, SL}.  Here
we give the proof of the second pair.
 
\medskip \noindent
{\it Proof of Eq.\re{funceqb}:}
We transform the
coefficients $\caA_{\uu', \uu''}^-$ of the expansion \re{ExpA} of
$\caA^-_\uu[f]$ as
\be\la{coefA} \caA_{\uu', \uu''}^-
= \prod _{ u''\in\uu''} {E^-
_{\uu'}(u'')} 
 = \prod _{ u''\in\uu''} {E^-
_{\uu'}(u'')\over E^+_{\uu'}(u'')}\ {E^+_{\uu'}(u'') }= ( - 1)^{\# \uu''}  \prod
_{u''\in \uu''} {E^-_{\uu}(u'')\over E^+_{\uu}(u'')}\  {E^+_{\uu'}(u'')
} \, ,  \ee
where we used the property $E^\pm_\uu=
E^\pm_{\uu'} E^\pm_{\uu''}$,  as well as the obvious identity
  \be \la{identity1} \prod _{u''\in \uu''} {E^- _{\uu''}(u'')\over
  E^+_{\uu''}(u'')}= ( - 1)^{\# \uu''}.  
  \ee
Summing  over all partitions, we obtain the expansion of the functional
$\caA^+[g]$, with $g = -E^-_\uu/E^+_\uu\, f$. $\square$

  \subsection{A  symmetric  expression for the inner product}

We can associate with the functional $\caA^\pm _\uu[f]$ a difference
operator $\hat \caA^\pm _\uu[f]$ acting on the functions on the set
$\uu$, by replacing $ f(u)\ \to\ f(u) \, e^{\pm i\p/\p u}$ for all
$u\in\uu$.  The operator $\hat \caA^\pm _\uu[f]$ is well defined if
the function $f(u)$ does not depend implicitly on the variables $\uu$.
The c-functional $\caA_\uu[f]$ is the result of the action of the
operator functional $\hat \caA^\pm_\uu[f] $ on the constant function
$1$, \be \no \caA^\pm _\uu[f]= \hat \caA^\pm _\uu[f] \cdot 1\, .  \ee
Then the expression [Eq.  \re{FactorFb}] for the inner product can be
laid down in terms of the functionals $\caA_\uu^-$ and $\hat
\caA_\vv^+$ as
 \be 
%\encadremath{ 
\la{ofacf} \caS_{\uu,\vv} =(-1)^{\# \uu}\ \hat
 \caA^+_\vv[ E^+_\uu/E^+_\zz]\cdot \caA^-_\uu[E^+_\vv]\, . 
 %} 
 \ee
Eq.  \re{ofacf} is equivalent to  the operator factorization
formula derived in \cite{3pf-prl, SL}.

Below we give an alternative expression for the inner product,
symmetric with respect to the sets of rapidities $\uu$ and $\vv$.
Define the functional
 \be
 \la{SLsym}
 \tilde \caS_{\uu,\vv} 
 &\defeq&
  (-1)^{\# \uu}\ \hat \caA^+_\vv[
 E^+_\uu/E^+_\zz]\cdot \caA^+_\uu[E^+_\vv/E^+_\zz]\, 
 \ =  \tilde \caS_{\vv,\uu}.
 \ee
 If the rapidities $\uu$ in Eq.  \re{ofacf} are assumed to be on
 shell, we can apply the first identity \re{funceqb} to transform the
 second factor on the rhs as
  \be \la{strgeqs}
   \caA^-_\uu[ E^+_\vv] \overset{\re{funceqb}  } 
   =
    \caA^+_\uu[ -  {E^-_\uu\over E^+_\uu}  E^+_\vv]
 \overset{\re{defquazimomentum}}
  = \caA^+_\uu[ -e^{2i p_\uu}  {E^+_\vv\over E^+_\zz}]
   \overset{\re{BAE} }= \caA^+_\uu[ { E^+_\vv\over E^+_\zz}] .
   \ee
Therefore we can identify
  \be
  \la{SLtSL}
   \caS_{\uu,\vv}= \tilde\caS_{\uu,\vv} \quad \text{for} \ \uu\
  \text{on shell}. \ee

The functional $ \tilde\caS_{\uu,\vv}$ can be simplified further.  It
is given, up to a sign, by the functional $\caA_\ww[1/E^+_\zz]$, with
$\ww=\uu\cup\vv$:
\be\la{tilSLA}
\tilde \caS_{\uu, \vv}=(-1)^{\# \uu}\ 
\caA^+_{\uu\cup\vv}[ 1/E^+_\zz] .
\ee
 {\it Proof:}
Using the definition \re{defCA}  and decomposing
\be \Delta_{\uu\cup\vv}= \Delta_\uu\, \ \Delta_\vv\prod _{u\in\uu,
v\in\vv} (u-v), \ee
  we write $\caA^+_{\uu\cup\vv}[ 1/E^+_\zz] $ as a product of two
  operators, one depending explicitly only on the set $\uu$, and the
  other depending on the set $\vv$: 
\be \la{2nx2ndet}
      \begin{aligned}
\caA^+_{\uu\cup\vv}[ 1/E^+_\zz] = & {1\over\Delta_\vv} \prod_{v\in\vv
} \( 1- {E^+_\uu(v)\over E^+_\zz(v) }\ e^{i\p/\p {v}} \) \,
\Delta_\vv\ \cdot {1\over\Delta_\uu} \prod_{u\in\uu}\( 1- {E^+_\vv (u)
\over E^+_\zz(u)}\ e^{i\p/\p {u}} \) \, \Delta_\uu\, \\
&= \hat \caA^+_\vv[E^+_\uu/E^+_\zz]\cdot \ \caA^+_\uu[ E^+_\vv/E^+_\zz]
\\
&=(-1)^{\# \uu}
\tilde \caS_{\uu, \vv}.
     \end{aligned}
     \ee

   As an example, apply this formula for the inner product 
   [Eq. \re{onemagnongen}] of    two one-magnon states:
     \be
     \begin{aligned}
    \tilde  \caS_{u,v}= \caA^+_{\{u, v\}}[1/E^+_\zz]
     =&
    1- {E^+_v(u)/ E^+_\zz(u)}
    -{E^+_u(v)/E^+_\zz(v) }
    +
    { 1 /E^+_\zz(u)  E^+_\zz(v)} 
      \no
    \\
    =&(1-{1/ E^+_\zz (u)} )(1-{1/E^+_\zz (v)})
    +  i\  { {1/ E^+_\zz (u)}  - {1/E^+_\zz (v)} \over u-v} .
       \end{aligned}
         \ee
If the rapidity $u$ is on shell, then is $ E^+_\zz =1$ and the first
term disappears, while the second gives the inner product.

A representation of the rhs of \re{tilSLA} as a Fock expectation value
of chiral fermions is given in the  Appendix.

\section{The  Slavnov determinant as a pDWPF}
\la{sec:SL-Z}
 
% \subsection{Gaudin-Izergin determinant, DWPF and  and 
% pDWPF}
%

 The inner product of two $M$-magnon states can be thought of as as a
 partition function of the six-vertex model on a $2M\times L$
 rectangular grid, with particular boundary conditions.  The
 $R$-matrix $R(u,v)$ [Eq.  \re{defR}] can be graphically represented
 as two intersecting segments, a horizontal one carrying a rapidity $u$
 and a vertical one carrying rapidity $v$,
\be \bR (u-v)= \setlength{\unitlength}{3pt} \begin{picture}(15, 9)(-2,5)
\put(-2.5, -1){\includegraphics[height=1.9cm]{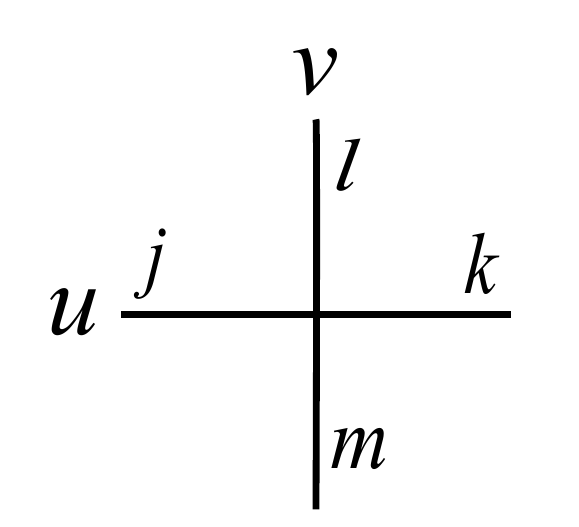}} \end{picture}
\quad = (u-v) \d_{jk}\d_{lm} + i\, \d_{jl}\d_{km}\, . \\ \no \ee
The two halves of each segment are labeled by $su(2)$ indices,   
which are represented in a standard way by arrows.  Then the  six
non-vanishing elements of the $R$-matrix correspond to the  
vertices of  the six-vertex model in the rational limit \cite{Baxter},
depicted in Fig.  \ref{fig:vertices}, with Boltzmann weights
 \be \la{6vweights} a(u,v) = u-v+i, \quad b(u,v) = { u-v}, \quad
 c(u,v) = { i}.  \ee
 The operators $\CB(u)$ and $\CC(u)$ are graphically represented in
 Fig.  \ref{fig:Monodromy}.  The  inner product [Eq.  \re{FockSL}]
 can be identified as the partition function of the six-vertex model
 defined on a $2M\times L$ rectangular grid, shown in Fig.
 \ref{fig:SLframe}.  The partition function is a sum over all possible
 ways to associate arrows with the internal links, so that at each
 site of the lattice the number of the incoming arrows equals the
 number of the outgoing arrows.  The boundary conditions on the arrow
 configurations are the following: on the two vertical boundaries, the
 lower half of the arrows point outwards, while the upper half point
 inwards.  On the two horizontal boundaries all arrows point
 upwards.  For example, the scalar product of two one-magnon states,
 Eq.  \re{onemagnongen}, is represented as
\be \la{oneM} \langle v|u\rangle =\sum_{k=1}^L
\setlength{\unitlength}{3pt} \begin{picture}(25, 15)(-4,5) \put(-3.5,
-4){\includegraphics[height=2cm]{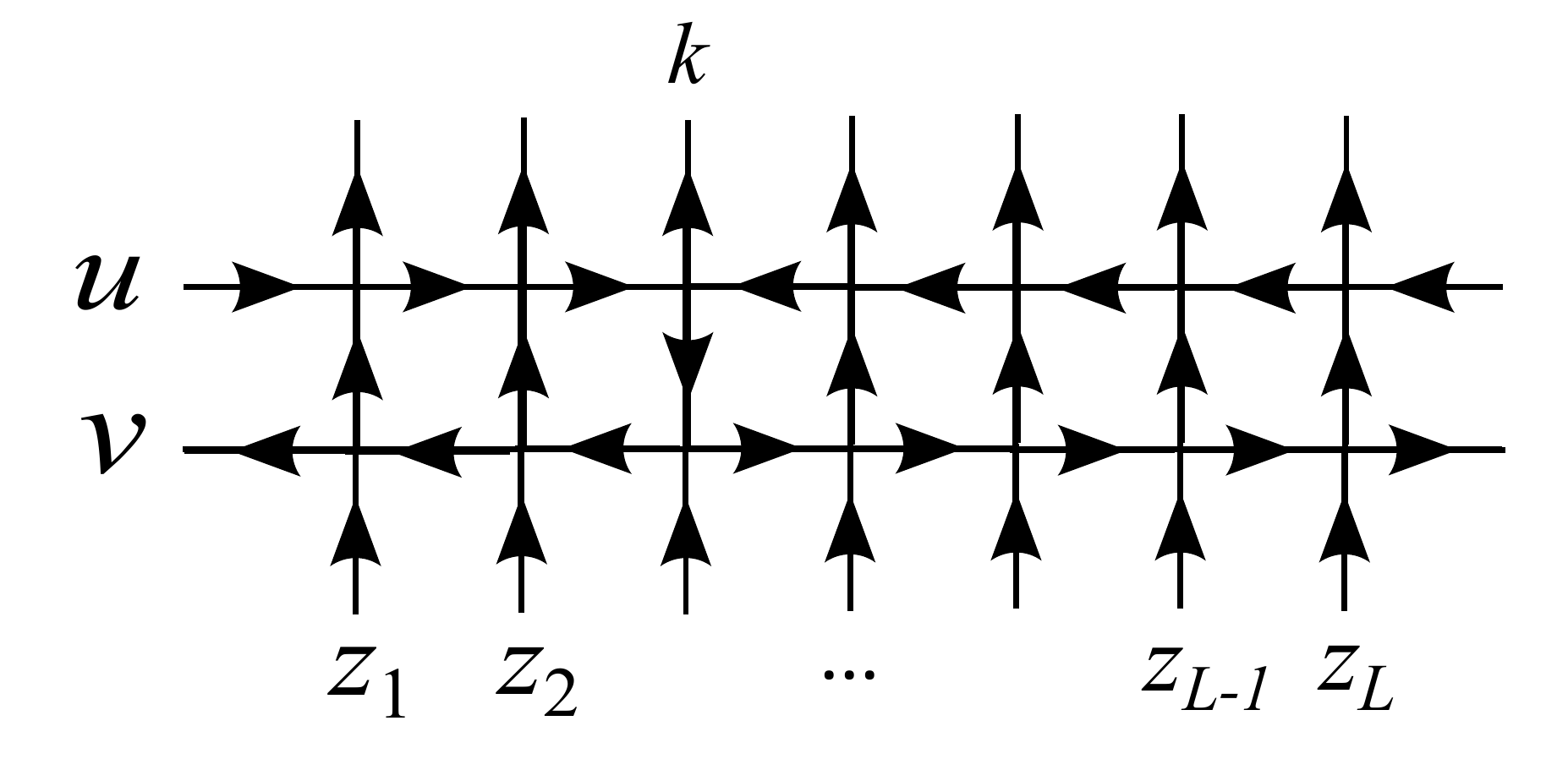}} \end{picture} \hskip 1.7cm
.  \ee
\vskip 0.8cm

\begin{figure}
         \centering
         %%----start of first figure----
	 \begin{minipage}[t]{0.4\linewidth}
            \centering
            \includegraphics[width=5.2 cm]{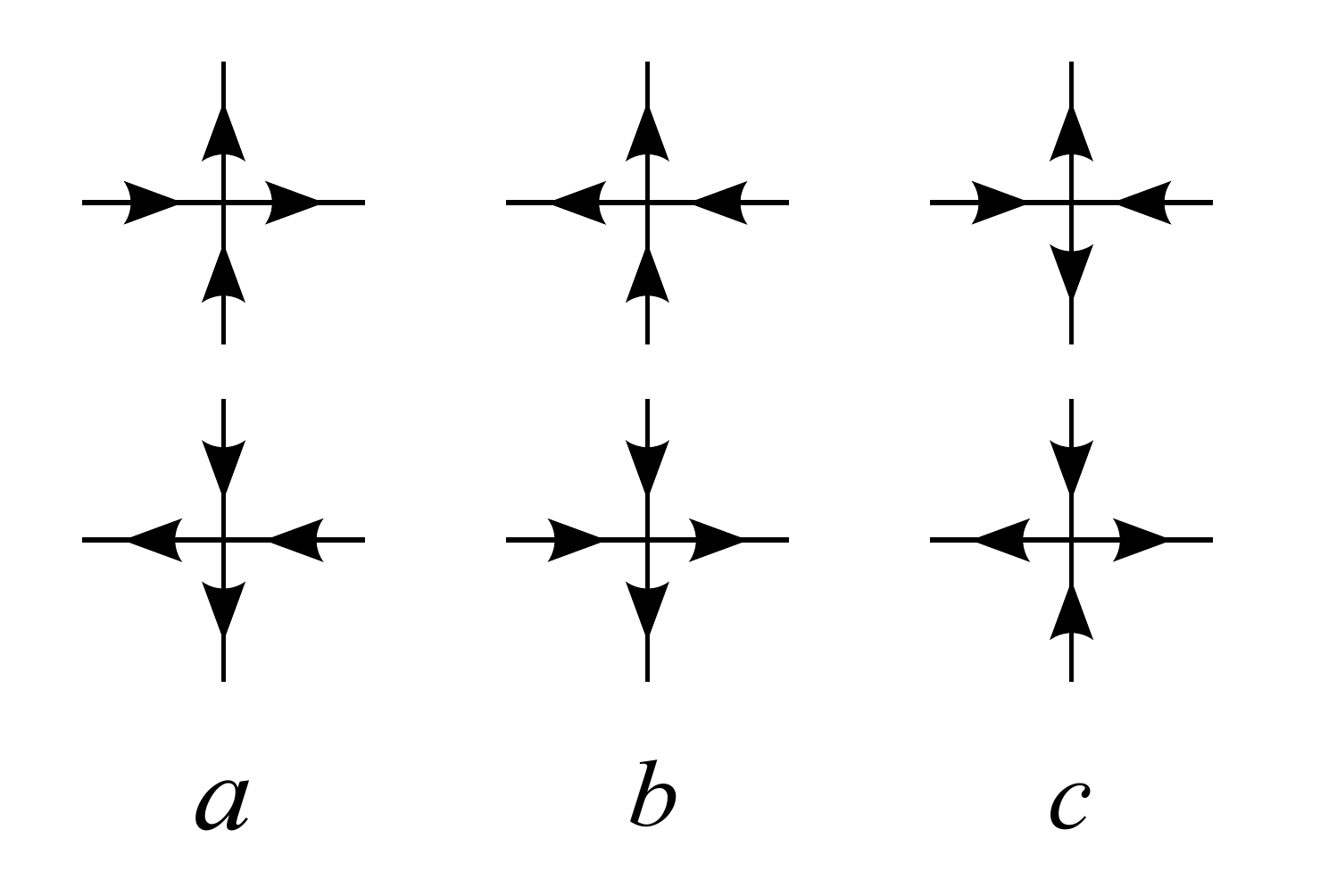}
  \caption{ \small Graphical representation of the
   non-vanishing elements of the $R$-matrix
  represented by the vertices of the six-vertex model.   }
  \label{fig:vertices}
         \end{minipage}%
         \hspace{2.5cm}%
         %%----start of second figure----
         \begin{minipage}[t]{0.4\linewidth}
            \centering
            \includegraphics[width=5.5cm]{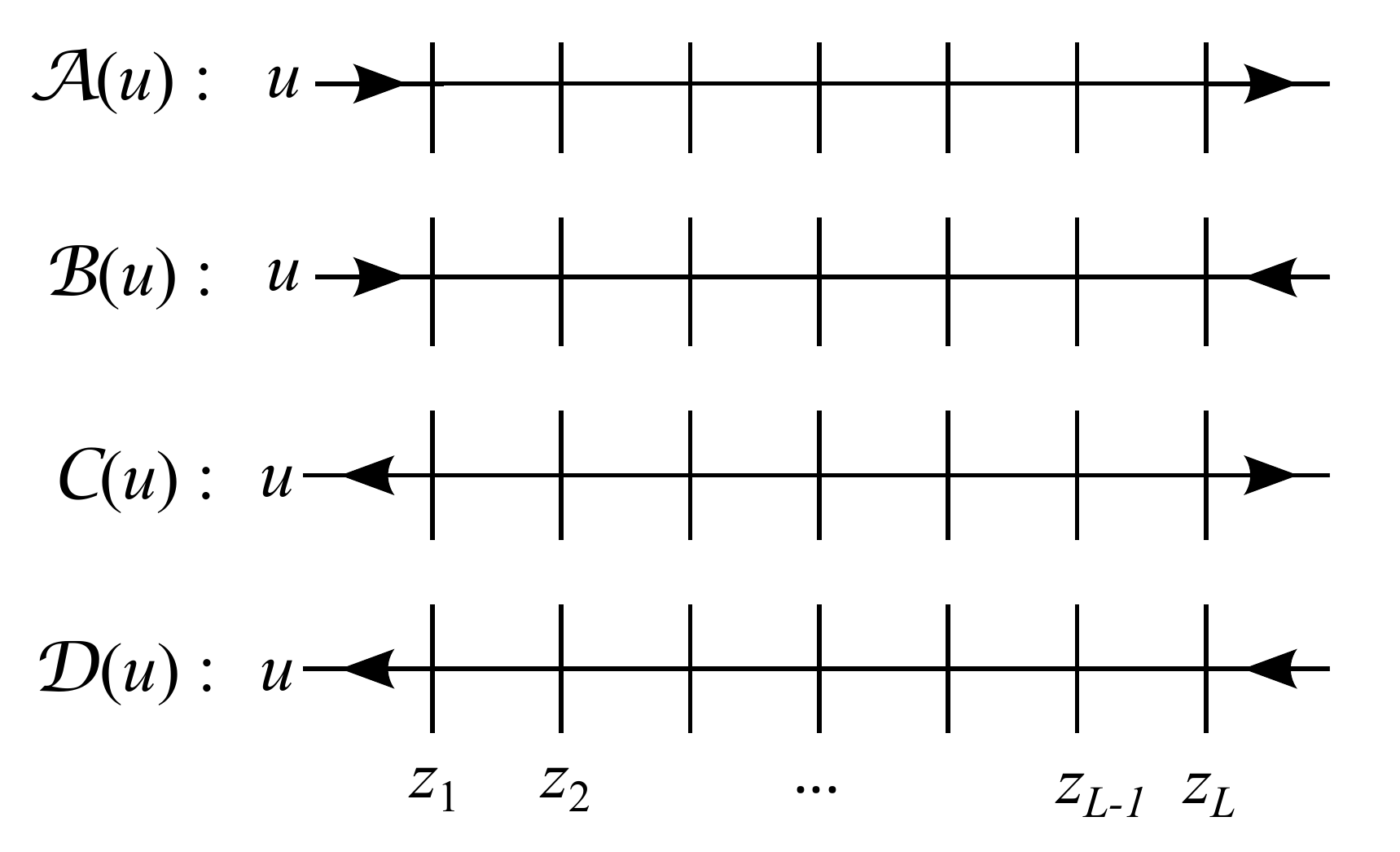}
	  \caption{ \small  Graphical representation of the
	   elements of the monodromy matrix}
\label{fig:Monodromy}
         \end{minipage}
           \end{figure}

 \begin{figure} 
%     \vskip 1cm
         %%----start of first figure----
	 \begin{minipage}[t]{0.4\linewidth}
            \centering
            \includegraphics[width=6.5 cm]{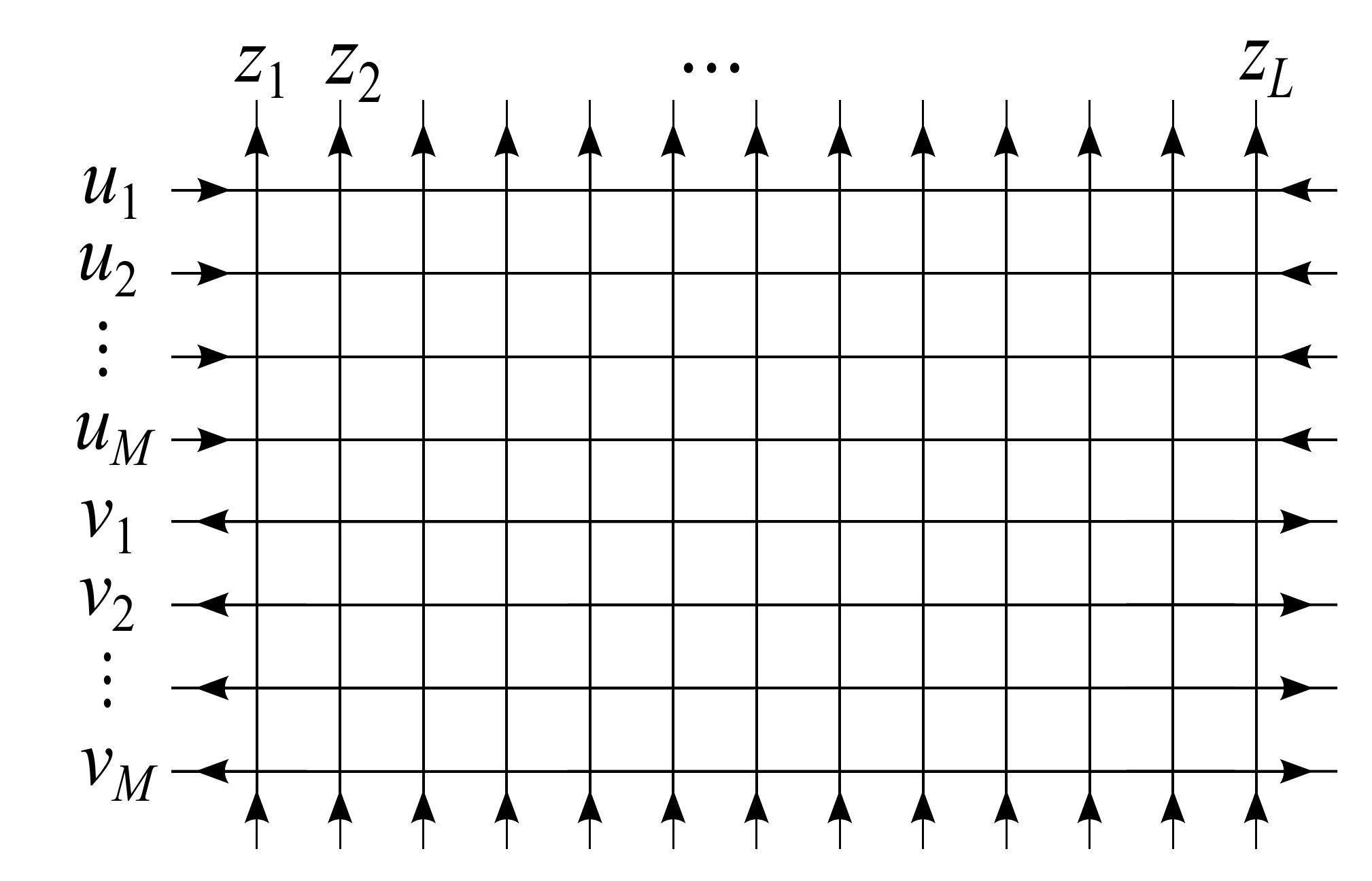}
\caption{ \small Graphical representation of the inner product
$\langle \vv|\uu\rangle$ as a six-vertex partition function on a
rectangular grid.  }
  \label{fig:SLframe}
         \end{minipage}%
         \hspace{1.8cm}%
                \begin{minipage}[t]{0.4\linewidth}
            \centering
            \includegraphics[width=4.3cm]{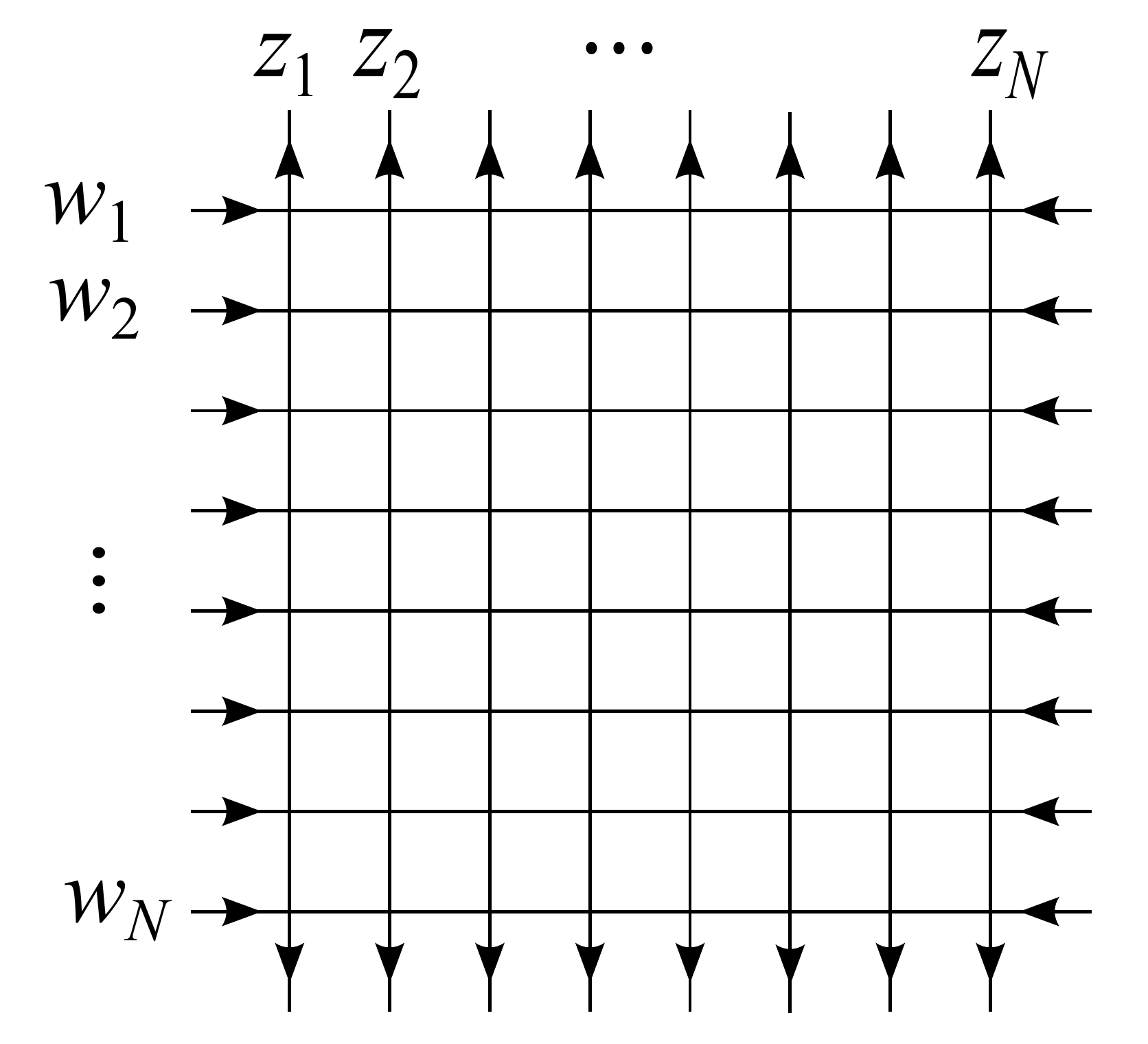}
	     \caption{ \small The domain wall boundary conditions
	     (DWBC).  }
\label{fig:DWBC}
         \end{minipage}
 \vskip1cm
         %%----start of first figure----
	 \begin{minipage}[t]{0.4\linewidth}
        %    \centering
            \includegraphics[width=6.5 cm]{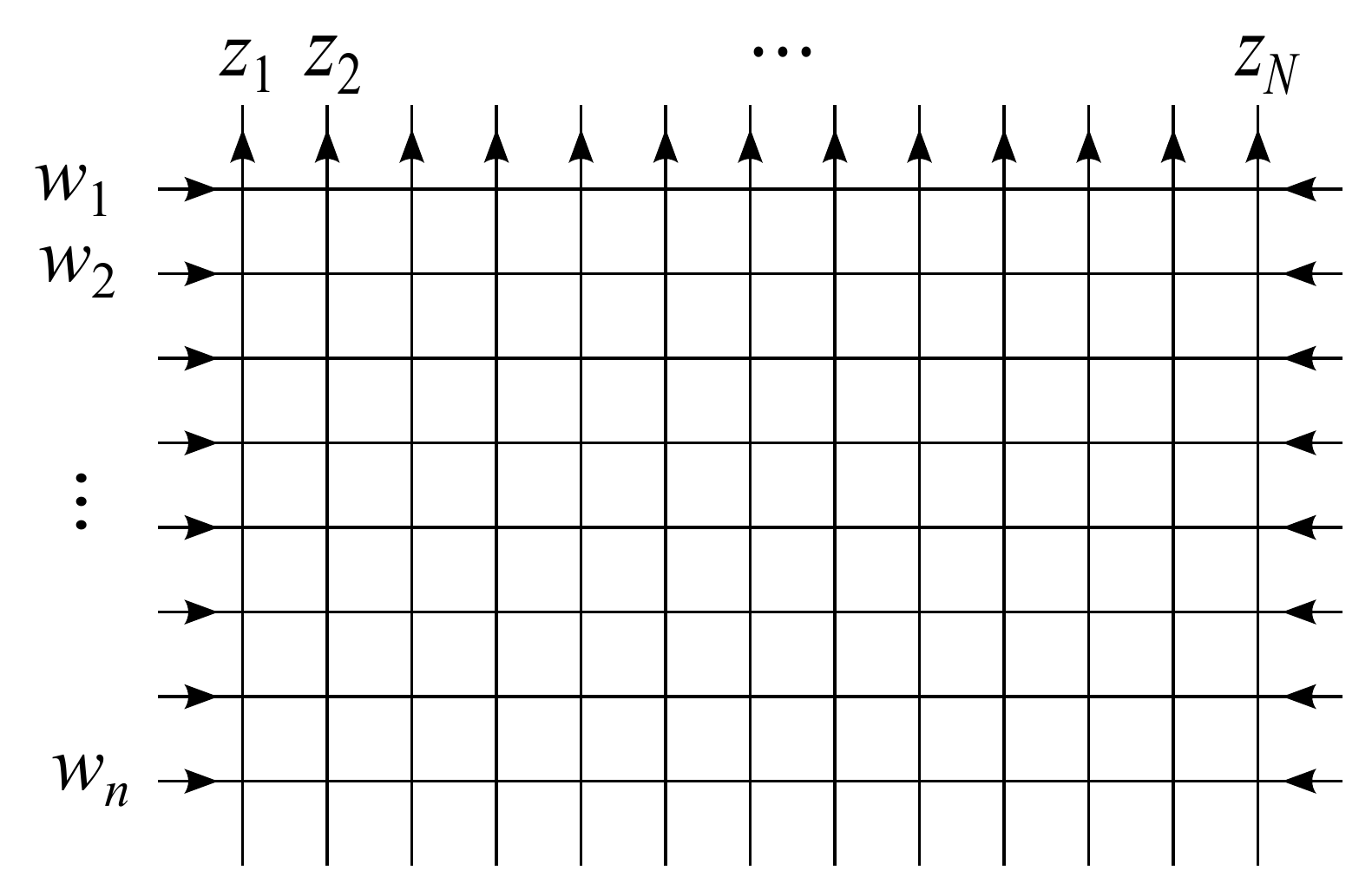}
\caption{ \small  The partial domain wall boundary conditions (pDWBC).
The boundary arrows on the top point outwards, those   on 
the vertical boundaries point inwards, and those
on the bottom segment  are free.}
  \label{fig:pDWBC}
         \end{minipage}%
         \hspace{1.8cm}%
                \begin{minipage}[t]{0.4\linewidth}
            \centering
            \includegraphics[width=6.9cm]{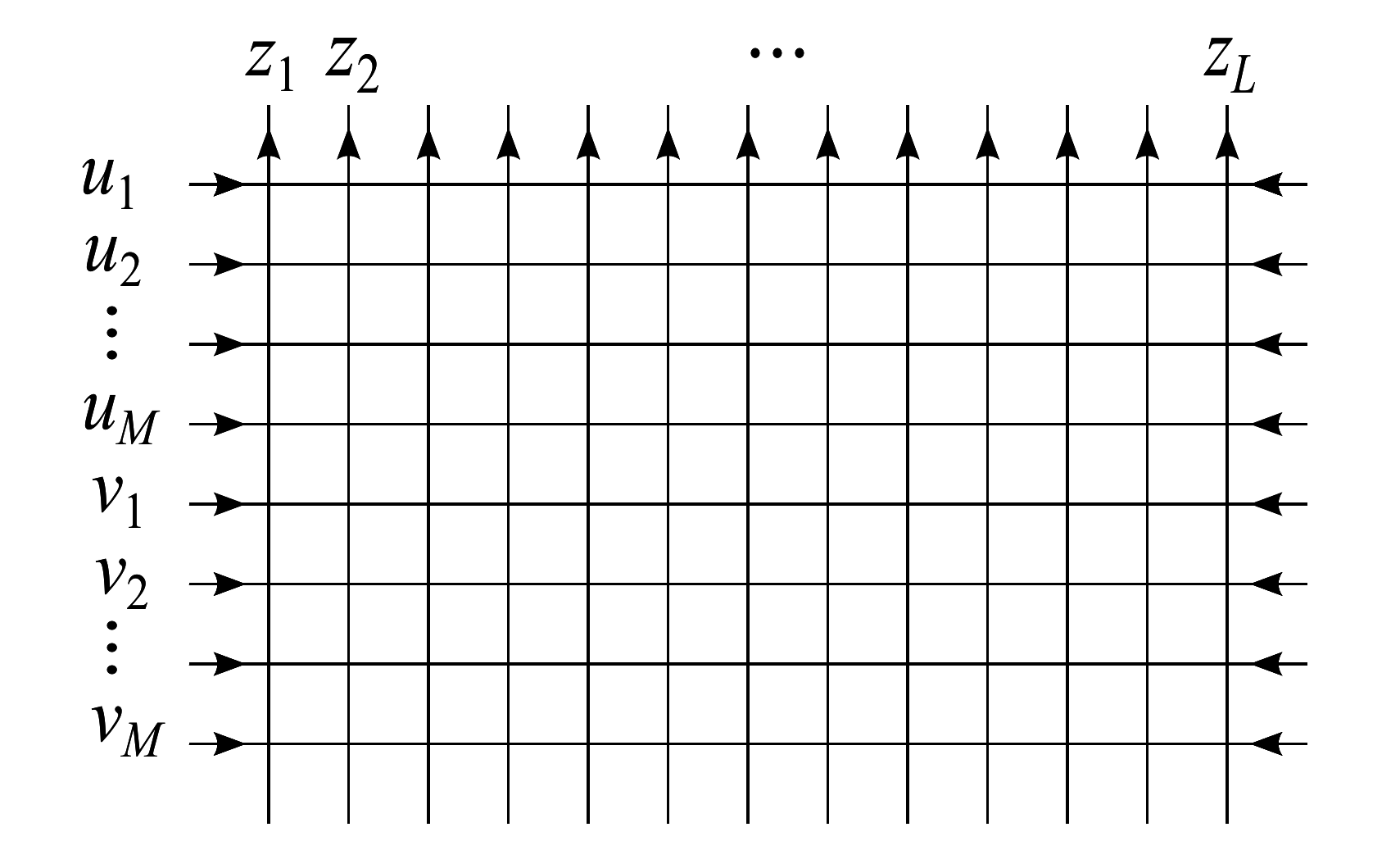}
	     \caption{ \small The pDWBC relevant for the Slavnov inner
	     product.  }
\label{fig:SL-pDW}
         \end{minipage}
% \vskip 1cm
      \end{figure}

%\vskip 0.8cm
The identity \re{SLtSL} means that the Slavnov inner product has a
second interpretation as a partition function of the six-vertex model.
Before formulating the correspondence, we give a brief recollection of
the (partial) domain wall partition functions.

The domain wall partition function, DWPF, is the partition function of
the six-vertex model on a square grid with domain wall boundary
conditions, DWBC \cite{korepin-DWBC,2000nlin......8030K}.  The DWBC
restrict the arrows on the edges forming the two horizontal boundaries
to point outwards, while the arrow on the edges forming the two
vertical boundaries point inwards, as shown in Fig.  \ref{fig:DWBC}.
The weights depend on the rapidity variables $\ww=\{w_1, \dots, w_N\}$
and $\zz= \{ z_1, \dots, z_N\}$, associated respectively with the
horizontal and with the vertical lines.  As shown by Izergin
\cite{Izergin-det,Izer-Cok-Kor}, the DWPF can be expressed, up to a
factor which can be eliminated by a renormalization of the six-vertex
weights, as a determinant\footnote{ For the first time the ratio of
determinants \re{IzergSP} appeared in the works of M. Gaudin
\cite{GaudinKorDet, Gaudin-livre} as the scalar product of two Bethe
wave functions for a Bose gas with point-like interaction on an
infinite line.  }

\be \la{IzergSP} \caZ_{\ww, \zz} &=& {\det_{j k} t(w_j - z_k ) \over
\det_{j k} {1\over w_j -z_k +i} } \, , \quad t(u) = {1\over u} -
{1\over u+i}\, . 
 \ee

 The Gaudin-Izergin determinant \re{IzergSP} is a particular case $M=
 L =N$, $\uu=\ww$, $\vv=\zz$, of the Slavnov determinant \re{defSuv}.
 Indeed, if $\vv=\zz$, the second term in the Slavnov matrix $\O$
 vanishes, $\O(u,v)=t(u-v)$.  Then the first factor in \re{FactorFb}
 is equal to 1, and we find \cite{SL}
  \be \la{IzerGauddet} \caZ_{\ww, \zz} =(-1)^{N}
  \caA^-_{\ww}[E^+_{\zz}] = (-1)^{N} \caA_\zz^+[E^-_\ww]\, , \quad
  \# \ww=\# \zz =N .  \ee
 Eq.  \re{IzerGauddet} can be derived also directly from the
 definition \re{IzergSP}, proceeding in the same way as we did in the
 case of the Slavnov determinant.  In this derivation one does not
 need to assume that the rapidities $\ww$ are on shell.

 The quantities named in \cite{FW3} {partial domain wall partition
 functions}, or pDWPF, are defined on a rectangular grid $n\times N$,
 where $1\le n\le N$.  The boundary conditions are the same as the
 DWBC except for the bottom segment of the boundary.  On the $N\times
 N$ square grid, the positions of the bottom arrows are fixed, once we
 have imposed the DWBC on the left, the right and the top segments of
 the boundary.  When $n<N$, this is no more the case.  The partial
 DWBC, depicted in Fig.  \ref{fig:pDWBC}, do not impose any
 restriction to the positions of the arrows on the bottom segment, and
 the bottom arrow configurations are summed over, just as those for
 the bulk segments.  We will denote the pDWPF again by $\caZ_{\ww,
 \zz}$, with $1\le \# \ww\le \# \zz $.  When this is needed by the
 context, we will denote by $\ww_n$ a set of rapidities $\ww$ with
 cardinality $\# \ww=n$.

 The pDWPF can be obtained from the DWPF by sending the rapidities
 $w_{n+1}, \dots, w_N$, associated with the bottom $N-n$ rows,
 sequentially to infinity.  Let $\uu_n$ be the subset of the first $n$
 rapidities in $\uu_N=\{u_1,\dots, u_N\}$.  Then the result of sending
 the remaining $N-n$ rapidities to infinity is \cite{FW3, SL}
  \be \la{pDWPFA} \caZ_{\uu_n, \zz_N} = \lim _{u_1\to\infty} {u_1\over
  i} \dots \lim _{u_N\to\infty} {u_N\over i} \caZ_{\uu_N , \zz_N} =
  (-1)^{n} (N-n)!\ \caA^-_{\uu_n }[E^+_{\zz_N}].  \ee

Applying the second identity \re{funceqa} to the rhs of \re{tilSLA},
we can relate the functional $\tilde \caS_{\uu, \vv}$ to the partial
domain wall partition function $\CZ_{\uu\cup \vv, \zz}$ defined on the
rectangular $2N\times L$ grid, as the one shown in Fig.
\ref{fig:SL-pDW}, where $N=\# \uu$ and $L=\# \zz $,
\be \la{Izerg-Sl} 
%\encadremath{ 
\tilde \caS_{\uu, \vv} = (-1)^{\# \uu}\
{ \caZ_{\uu\cup\vv,\, \zz}\over \CN_{\uu\cup \vv,\, \zz}}\ \ \, \ ,
\qquad \CN_{\ww, \, \zz} \defeq (\# \zz - \# \ww)!  \prod_{w\in\ww}
E_\zz^+(w)\, .
%} 
\ee
Another way to write the proportionality factor is as
\be \CN_{\ww,\ \zz} =(\#\zz-\# \ww)!\ \prod_{w\in\ww} {A(w)\over
D(w)}.  \ee

The proof Eq.  \re{Izerg-Sl} is surprisingly simple.  First we
transform the representation \re{pDWPFA}, using the first of the
functional identities \re{funceqa},
\be \la{ZuvA} \caZ_{\uu\cup \vv, \zz}= ( L- 2M)!  \
\caA^-_{\uu\cup\vv}[ E^+_\zz] ={ \CN_{\uu\cup \vv,\, \zz}}\ \
\caA^+_{\uu\cup\vv}[ 1/E^+_\zz] .  \ee
This concludes the proof of  \re{Izerg-Sl}.

\section{Concluding remarks}

In this paper, we derived an expression for the inner product of an
$M$-magnon Bethe eigenstate and an $M$-magnon generic state in the
inhomogeneous periodic XXX chain of length $\#\zz= L$, which is
completely symmetric in the union of the two sets of rapidity
parameters:
\be
\la{2nx2ndetA}
      \begin{aligned}\tilde \caS_{\uu, \vv}
&=(-1)^{\# \uu} \caA^+_{\uu\cup\vv}[ 1/E^+_\zz] .
     \end{aligned}
     \ee
The functional $\tilde \caS_{\uu, \vv}$ represents a determinant
$2M\times 2M$ and is given essentially by the partition function with
domain boundary conditions on a $L\times L$ square grid, with $L-2M$
of rapidities sent to infinity.  The functional $\tilde \caS_{\uu,
\vv}$ coincides with the original Slavnov product \re{ofacf} if the
rapidities $\uu$ are on shell, and for periodic boundary conditions
(no twist in the Bethe equations).  In general, $\tilde \caS_{\uu,
\vv}$ and $ \caS_{\uu, \vv}$ are two distinct functionals.

An immediate application of the representation \re{2nx2ndetA} in the
study of the semiclassical limit of the three-point function of long
trace operators in the $su(2)$ sector $\CN=4$ SYM, formulated in Refs.
\cite{EGSV,GSV}.  Using the determinant representation given in 
 \cite{Omar}, a closed expression for the structure constant for
three non-protected operators was obtained in Refs.  \cite{3pf-prl,
SL} as a generalization of the result for one-protected and two
non-protected operators operators found in Ref.  \cite{GSV}.  The
classical limit of the functionals $\caS_{\uu, \vv}$ and $\tilde
\caS_{\uu, \vv}$ is the same, but if one is interested in the
subleading terms, second functional is much more convenient to deal
with.  On the other hand, the expression \re{2nx2ndetA} with generic
inhomogeneity parameters can be used to reproduce the higher orders in
the weak coupling expansion of the structure constant, as it has been
argued in \cite{2012arXiv1202.4103G, Didina-Dunkl}.

The alternative representation of the inner product \re{FockSL}, found
in this paper, has a natural interpretation on terms of the Fock space
for the Algebraic Bethe Ansatz.  The functional $\tilde \caS_{\uu,
\vv}$ is proportional to the inner product
 \be \tilde \caS _{\uu, \vv} &\sim & \bradown{L}\, (
 \bS^-)^{L-2M}\prod_{j =1}^M \CB(v_j ) \prod_{j=1}^{M} \CB(u_j )
 \ketup{L}\,\no \\
& \sim&
 \braup{L}\, \prod_{j =1}^M
 \CC(v_j ) \prod_{j=1}^{M} \CC(u_j )
\ ( \bS^+)^{L-2M} \ketdown{L} \, .
 \la{algebrt}
 \ee
The second functional has the same structure as the result of a
particle-hole transformation on the ket vector on the rhs of the
original inner product [Eq.  \re{FockSL}].  In case of a non-zero
twist, the dual rapidities are different than the original rapidities
\cite{Bazhanov:2010aa}.  In the case we are considering, $M$ of the
dual rapidities coincide with the original ones, while the the rest
$L-2M$ of them go to infinity.  This can be justified by the following
simple argument.\footnote{We thank D. Serban for suggesting to us this
argument.} For zero twist, the global $su(2)$ symmetry is not broken
and the Bethe states is a direct sum of states with given spin, which
are eigenspaces of the transfer matrix.  The states corresponding to
the same solution of the Bethe equations, belong to the same $su(2)$
multiplet \cite{Faddeev}.  Therefore the Bethe eigenstates $\prod
_{u\in \uu} \CB(u)\ketup{L}$ and $\prod _{u\in \uu} \CC(u)\ketdown{L}$
must be related by the action of a global raising operator, which must
be $(\bS^+)^{L-2M}$, since the first state has $S^z = \hf L-M$, while
the second state has $S^z = M-\hf L$.  Thus we have
 \be
 \prod_{j=1}^{M} \CC(u_j )
\ ( \bS^+)^{L-2M} \ketdown{L} \,
\ \sim\  \prod_{j=1}^{M} \CB(u_j )\
\ketup{L}\, ,
\ee
where the proportionality sign means that the two states are equal up
a c-function of the rapidities.  In this sense, our main result is the
computation of the factor of proportionality.

\bigskip

\noindent {\it Note added.} After the preparation of the manuscript,
we learned about the publication \cite{2012arXiv1207.2352F}, 
which is along the same lines as our work.

  \section*{Acknowledgments}
  
I.K. thanks O. Foda, D. Serban, A. Sever and F. Smirnov for useful discussions.
YM is grateful to the hospitality of the collegues in Saclay during
his stay.  This research has received funding from the [European Union] Seventh Framework Programme [FP7-People-2010-IRSES]  
under grant agreement N$^o$  269217,  the PHC SAKURA 2012,
Projet N$^o$ 27588UASakura and corresponding Grant from Japan.  YM is
partially supported by Grant-in-Aid (KAKENHI \#20540253) from MEXT
Japan.

\appendix

   \section{Fermionic representation of the functionals $\caA^\pm_\uu[f]$}
   }

 The functionals $\caA^- _\uu[E^+_\zz] $ and $\caA^+_\uu[1/ E^+_\zz]$
 have natural fermionic representations.  Introduce a chiral
 Neveu-Schwarz fermion living in the rapidity complex plane and having
 mode expansion
\be \la{pzpo} \psi (u)= \sum_{r\in \IZ+ {1\over 2}}\psi_{r}\ u^{-r-
{1\over 2}}, \ \ \ \ \bar \psi (u)= \sum_{r\in \IZ+ {1\over 2}} \bar
\psi_{ r}\ u^{r- {1\over 2}} .  \ee
The fermion modes are assumed to satisfy the anticommutation relations
\be\la{cpmto} [\bar \psi _{ r}, \bar \psi _{ r'} ]_+= [ \psi _{ r},
\psi _{ r'} ]_+= 0\, , \quad [ \psi _{ r}, \bar \psi _{ r'} ]_+=
\delta_{r, r'}\, , \ee
and the left/right vacuum states are defined by
\be\la{mnfio2} \bra \psi_{-r}= \bra \bar \psi_{r} = 0\ \ \text{and}\ \
\ \psi_{r}\, \ket =\bar \psi_{-r} \ket = 0,\ \ \ \text{for} \ r> 0 .
\ee
The operator $\bar \psi_r$ creates a particle (or annihilates a hole)
with mode number $r$ and the operator $\psi_r$ annihilates a particle
(or creates a hole) with mode number $r$.  The particles carry charge
1, while the holes carry charge $-1$.  The charge zero vacuum states
\re{mnfio2} are obtained by filling the Dirac sea up to level zero.
The left vacuum states with integer charge $\pm N$ are constructed as
 \be
 \la{defchargeN}
 \langle N|\ = \ 
 \begin{cases}
   \bra \psi_{1\over 2} \dots \psi_{N-{1\over 2}} & \text{ if}\ N>0 ,
   \\
  \bra \bar \psi_{-{1\over 2}}\dots \bar\psi_{-N+{1\over 2}} &
  \text{if} \ N<0 .
\end{cases}
   \ee
 Any correlation function of the operators \re{pzpo} is a determinant
 of two-point correlators
\be \la{opepsi} \bra \psi(u) \bar \psi(v)\ket = {1\over u-v}\, .  \ee

 The following formulae is  easily established.
 Let $\# \uu=n $ and $\# \zz = N$. Then
\be
\la{fermionicA}
\begin{aligned}
\caA^+_\uu[ 1/E^+_\zz]&=
 {\langle N-n|  \prod_{j =1}^n \[ \psi(u_j ) - \psi(u_j
+i) \]
 \ \prod_{k =1}^N \bar\psi(z_k ) \ket \over\langle N-n| \prod_{j =1}^n
 \psi(u_j ) \prod_{k =1}^N\bar \psi(z_k )\ket } ,
\end{aligned}
\ee
\be
\la{fermionicB}
\begin{aligned}
\caA^-_\uu[ E^+_\zz]&=
{\langle N-n|  \prod_{j =1}^n \[ \psi(u_j ) - \psi(u_j
+i) \]
 \ \prod_{k =1}^N \bar\psi(z_k ) \ket \over\langle N-n|  \prod_{j =1}^n
 \psi(u_j +i ) \prod_{k =1}^N\bar \psi(z_k )\ket } 
.
\end{aligned}
\ee

Eq.  \re{fermionicA} gives a convenient representation of the pDWPF
and the Slavnov scalar product in terms of free chiral fermions.
Other fermionic representations have been proposed in
\cite{JPSJ.62.1887,FWZ,FoSh} for the Slavnov product and in \cite{
PZinn-review6v, CPZinn-2010, 2012arXiv1203.5621F} for the
Gaudin-Izergin determinant, see also the review paper
\cite{2010arXiv1003.3071T}.

%
%
% \footnotesize
%% 
% \bibliography{/Users/vani/Files/PAPERS/PAPERSLIBRARY/ABib}
%  \bibliographystyle{/Users/vani/Files/PAPERS/PAPERSLIBRARY/utcaps}
%

\providecommand{\href}[2]{#2}\begingroup\raggedright\endgroup

 \end{document}